\documentclass[a4paper,fleqn,usenatbib]{mnras}

\usepackage[T1]{fontenc}
\usepackage{ae,aecompl}

\usepackage{graphicx}	% Including figure files
\usepackage{amsmath}	% Advanced maths commands
\usepackage{amssymb}	% Extra maths symbols
\usepackage{graphics}
\usepackage{epsfig}
\usepackage{txfonts}
\usepackage{flushend}
\usepackage{blindtext}
\title[]{The luminous, massive and solar metallicity galaxy hosting the \\ \textit{Swift} $\gamma$-ray burst, GRB\,160804A at $z = 0.737$ \thanks{Based on observations carried out under the programme ID 097.A-0036 (PI: J. Fynbo) with the X-shooter spectrograph mounted at the Cassegrain Very Large Telescope (VLT), Unit 2 - Kueyen, operated by the European Southern Observatory (ESO) on Cerro Paranal, Chile; and on observations made with the Italian Telescopio Nazionale Galileo (TNG) operated on the island of La Palma by the Fundaci\'on Galileo Galilei of the INAF (Istituto Nazionale di Astrofisica) at the Spanish Observatorio del Roque de los Muchachos of the Instituto de Astrof\'isica de Canarias under program A32TAC\_5 (PI: D'Elia).}} % 

\author[K. E. Heintz et al.]{K.~E.~Heintz,$^{1,2}$ \thanks{E-mail: keh14@hi.is}
D.~Malesani,$^{2}$
K.~Wiersema,$^{3}$
P.~Jakobsson,$^{1}$
J.~P.~U.~Fynbo,$^{4}$
\newauthor
S.~Savaglio,$^{5}$
Z.~Cano,$^{6}$
S.~Covino,$^{7}$
V.~D'Elia,$^{8,9}$
A.~Gomboc,$^{10}$
F.~Hammer,$^{11}$
L.~Kaper,$^{12}$
\newauthor
B.~Milvang-Jensen,$^{2}$
P.~M\o ller,$^{13}$
S.~Piranomonte,$^{14}$
J.~Selsing,$^{2}$
N.~H.~P.~Rhodin,$^{2}$
\newauthor 
N.~R.~Tanvir,$^{3}$
C.~C.~Th\"one,$^{6}$
A.~de~Ugarte~Postigo,$^{6,2}$
S.~D.~Vergani,$^{11}$ 
\& D.~Watson$^{4}$
\\
$^{1}$Centre for Astrophysics and Cosmology, Science Institute, University of Iceland, Dunhagi 5, 107 Reykjav\'ik, Iceland\\
$^{2}$Dark Cosmology Centre, Niels Bohr Institute, University of Copenhagen, Juliane Maries Vej 30, 2100 Copenhagen \O, Denmark\\
$^{3}$Department of Physics \& Astronomy and Leicester Institute of Space \& Earth Observation, University of Leicester, University Road, \\ Leicester LE1 7RH, United Kingdom\\
$^{4}$Niels Bohr Institute, University of Copenhagen, Juliane Maries Vej 30, 2100 Copenhagen \O, Denmark\\
$^{5}$Physics Department, University of Calabria, via P. Bucci, I-87036 Arcavacata di Rende, Italy \\
$^{6}$Instituto de Astrof\'isica de Andaluc\'ia (IAA-CSIC), Glorieta de la Astronom\'ia s/n, E-18008, Granada, Spain\\
$^{7}$INAF - Osservatorio Astronomico di Brera, via Bianchi 46, 23807, Merate (LC), Italy\\ 
$^{8}$INAF - Osservatorio Astronomico di Roma, Via Frascati 33, I-00040 Monteporzio Catone, Italy\\
$^{9}$ASI-Science Data Centre, Via del Politecnico snc, I-00133 Rome, Italy\\
$^{10}$Centre for Astrophysics and Cosmology, University of Nova Gorica, Vipavska 11c, 5270 Ajdov\v s\v cina, Slovenia\\
$^{11}$GEPI, Observatoire de Paris, PSL Research University, CNRS, Place Jules Janssen, 92190 Meudon \\
$^{12}$Astronomical Institute Anton Pannekoek, Universiteit van Amsterdam, Postbus 94249, 1090 GE Amsterdam, The Netherlands  \\
$^{13}$European Southern Observatory, Karl-Schwarzschildstrasse 2, D-85748 Garching bei M\"unchen, Germany\\
$^{14}$INAF - Osservatorio Astronomico di Roma, Monte Porzio Catone (RM), 00078, Italy \\
}

\date{Accepted 2017. Received 2017; in original form 2017}

\pubyear{2017}

\begin{document}
\label{firstpage}
\pagerange{\pageref{firstpage}--\pageref{lastpage}}
\maketitle

\begin{abstract}
We here present the spectroscopic follow-up observations with VLT/X-shooter of the \textit{Swift} long-duration gamma-ray burst GRB\,160804A at $z = 0.737$. Typically, GRBs are found in low-mass, metal-poor galaxies which constitute the sub-luminous population of star-forming galaxies. For the host galaxy of the GRB presented here we derive a stellar mass of $\log (M_*/M_{\odot})$ = $9.80\pm 0.07$, a roughly solar metallicity ($12+\log(\mathrm{O/H}) = 8.74\pm 0.12$) based on emission line diagnostics, and an infrared luminosity of $M_{3.6/(1+z)}=-21.94$ mag, but find it to be dust-poor ($E(B-V)<0.05$ mag). This establishes the galaxy hosting GRB\,160804A as one of the most luminous, massive and metal-rich GRB hosts at $z<1.5$. 
Furthermore, the gas-phase metallicity is found to be representative of the physical conditions of the gas close to the explosion site of the burst. 
The high metallicity of the host galaxy is also observed in absorption, where we detect several strong Fe\,\textsc{ii} transitions as well as Mg\,\textsc{ii} and Mg\,\textsc{i}. While host galaxy absorption features are common in GRB afterglow spectra, we detect absorption from strong metal lines directly in the host continuum (at a time when the afterglow was contributing to $<15\%$). Finally, we discuss the possibility that the geometry and state of the absorbing and emitting gas is indicative of a galactic scale outflow expelled at the final stage of two merging galaxies. 
\end{abstract}


%%%%%%%%%%%%%%%%%%%%%%%%%%%%%%%%%%%%%%%%%%%%%%%%%%

%%%%%%%%%%%%%%%%% BODY OF PAPER %%%%%%%%%%%%%%%%%%

\section{Introduction}

It has now been firmly established that long-duration gamma-ray bursts (GRBs) are associated with the deaths of massive stars \citep{Woosley06a,Cano17} and should therefore also be connected to cosmic star formation \citep{Wijers98,Christensen04,Jakobsson05a,Kistler08,Robertson12,Greiner15}. Galaxies hosting GRBs can be studied at high redshifts \citep[e.g. at $z>5$;][within the first Gyr after the Big Bang]{Tanvir12,Basa12,Salvaterra13,Sparre14,Hartoog15,McGuire16} and even at very faint magnitudes \citep{LeFloch03,Savaglio09}. Galaxies that are intrinsically faint or at high redshifts constitute some of the major observational challenges in conventional luminosity-selected star-forming galaxy surveys and are therefore underrepresented in such samples. The study of GRB host galaxies is thus a valuable, complementary approach to probe the overall population of star-forming galaxies.

To reconcile the host galaxy population of GRBs to that of the general population of star-forming galaxies it is important to understand how the physical properties such as star-formation rate, stellar mass and metallicity influence the GRB production rate. Specifically at low redshifts ($z < 1.5$), GRBs have been shown to occur preferentially in low-metallicity environments \citep{Kruehler15,Schulze15,Japelj16,Vergani17}, translating into generally lower stellar masses and fainter luminosities for their host galaxies \citep{Sollerman05,Wolf07,Vergani15,Perley13,Perley16b}. It is now evident, however, that luminous, massive and hence metal-rich GRB host galaxies do exist but they are often associated with dusty or "dark" GRB afterglows \citep{Kruehler11,Perley13,Perley16b,Hunt14} and are therefore as a consequence underrepresented in samples selected by optical afterglow identification. While most single-star progenitor models require the GRB environment to be metal-poor \citep{Yoon06,Woosley06b}, super-solar metallicities in GRB hosts are observed \citep[as well as in afterglows, see e.g.][]{Savaglio12}, possibly due to significant internal chemical inhomogeneity within the hosts. This scenario is also supported by numerical simulations or semi-analytic models \citep{Nuza07,Niino11,Trenti15,Bignone17}, which show that even though metal-poor GRB host environments dominate the overall population, this model does not exclude a small number of near-solar metallicity hosts. However, the shape of the overall distribution of oxygen abundance in integrated host galaxy spectra can capture some of the statistical properties of GRB progenitors, making studies of the extreme ends of this distribution particularly important. This paper deals with such a case.

We present spectroscopic observations of the long-duration GRB\,160804A. GRB\,160804A was detected by \textit{Swift} on 2016 August 4, 01:32:47 UT \citep[][GCN 19761]{Marshall16}, with XRT observations starting 147 s after the BAT trigger. 
The burst had a fairly long $T_{90}$ duration of $144.2\pm 19.2$ s and a small best-fit absorption column density of $N_{\rm H,X} = 7.1^{+4.3}_{-3.8}\times 10^{20}$ cm$^{-2}$ in excess of the Galactic value\footnote{The \textit{Swift}-XRT repository can be found at:\\ \url{http://www.swift.ac.uk/xrt_spectra/}} \citep{Willingale13}.
We observed GRB\,160804A as part of the X-shooter GRB (XS-GRB) afterglow legacy survey (Selsing et al., in preparation, PI: J. Fynbo).

We study both the optical/near-infrared emission and absorption line properties of this GRB host galaxy in detail and have structured the paper as follows. In Sect.~\ref{sec:obs} we describe our observations and in Sects.~\ref{sec:hostem} and \ref{sec:abs} we present the results separately in emission and absorption. In Sect.~\ref{sec:conc} we summarize and conclude on our work.
Throughout the paper, we assume a flat concordance cosmological model with $H_0 = 67.8$\,km\,s$^{-1}$\,Mpc$^{-1}$, $\Omega_m = 0.308$ and $\Omega_{\Lambda}=0.692$ \citep{Planck16}. Unless otherwise stated all magnitudes are given in the AB \citep{Oke74} magnitude system and we use the photospheric solar abundances from \cite{Asplund09}.

\section{Observations and data reduction} \label{sec:obs}

\subsection{X-shooter spectroscopy}

We observed the optical/near-infrared afterglow of GRB\,160804A with the ESO/VLT Unit Telescope 2 (UT2, Kueyen) equipped with the X-shooter spectrograph \citep{Vernet11}. Observations started at 23:55 UT on 2016-08-04 (22.37 hr after the BAT trigger) and consisted of four spectra of 600\,s, observed following an ABBA nodding pattern, covering the wavelength range 3200~--~20000\,\AA.

The spectra were taken under good conditions with a median seeing of $0\farcs75$ at 6700\,\AA. The airmasses at the start and end of the spectroscopic observations were 1.30 and 1.45. The spectra were acquired using slit widths of $1\farcs0$, $0\farcs9$ and $0\farcs9$ for the UVB, VIS and NIR arm, respectively, approximately aligned with the parallactic angle. For the NIR arm observations we used a $K$-band blocking filter. For this given setup, the nominal instrumental resolution is 4290, 7410 and 5410\footnote{Table 2 at \url{https://www.eso.org/sci/facilities/paranal/instruments/xshooter/inst.html}.} for the UVB, VIS and NIR arm, respectively. Since the seeing was smaller than the slit widths used, the true resolution is higher than this. In the VIS and NIR arm we measure the resolution from the width of several atmospheric absorption lines and find it to be 28.3 km s$^{-1}$ ($\mathcal{R}_{\mathrm{VIS}}=10600$) and 37.5 km s$^{-1}$ ($\mathcal{R}_{\mathrm{NIR}}=8000$).

The X-shooter data reduction was done as part of the XS-GRB legacy sample and is described in detail in Selsing et al. (in prep.). The measured resolution in the VIS arm is perfectly consistent with the empirical relation studied by Selsing et al., tying the spectral resolution with the spatial extent of the trace and thus supports the use of this simple diagnostic tool. The ratio between the observed and the nominal resolution in the VIS arm is used to extrapolate the spectral resolution to the UVB arm (where no telluric lines are present) and we estimate a seeing-corrected resolution of 49.3 km s$^{-1}$ ($\mathcal{R}_{\mathrm{UVB}}=6090$).

\begin{figure*} %[ht!]
	\centering
	\epsfig{file=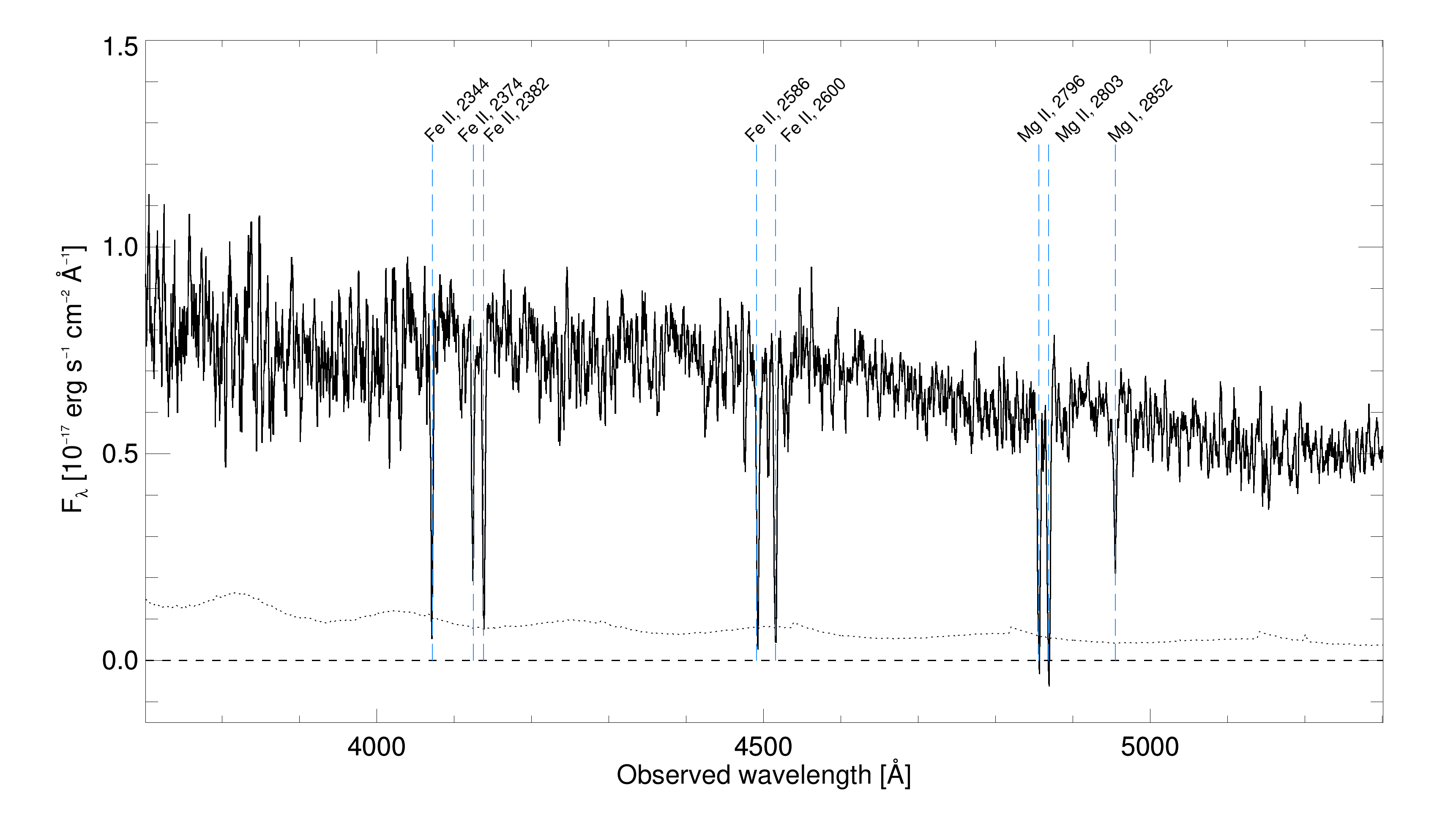,width=0.9\textwidth}
	\caption{The reduced 1D X-shooter spectrum (UVB arm) of GRB\,160804A shown as the black solid line. The spectrum has been smoothed to enhance details. The dotted line denotes the error spectrum. The strong, prominent absorption lines are each marked and can clearly be identified in the spectrum even in the absence of a bright afterglow due to their large equivalent widths.  \label{fig:uvbnorm}}
\end{figure*}

The spectrum revealed strong absorption features from Fe\,\textsc{ii}, Mg\,\textsc{ii} and Mg\,\textsc{i} (see Fig.~\ref{fig:uvbnorm}) and a spatially resolved H$\alpha$ line \citep{Xu16}. The absorption lines are clearly detected even though the continuum is dominated by the host rather than the afterglow ($\lesssim 15\%$ afterglow contribution at the time of observation, see below). Moreover, the absorption lines were observed to be at a common redshift ($z=0.737$) of the emission lines, ruling out an intervening absorber unrelated to the GRB host as the absorbing system. Galaxies can indeed show strong absorption features associated with a cold interstellar medium (ISM) if the host galaxy has high metallicity and/or if the total gas column density is high. However, GRB host spectra (as opposed to afterglow spectra) rarely show such strong absorption lines. This motivated our further investigation of this object. 

\subsection{Imaging} \label{ssec:ima}

The afterglow of GRB\,160804A was monitored by several imaging campaigns and we report the collected measurements from optical to near-infrared imaging in Table~\ref{tab:photdata}. The host galaxy of the GRB is also detected in the archival data of the Sloan Digital Sky Survey data release 13 \citep[SDSS-DR13, optical $u,g,r,i,z$-filters;][]{SDSSDR132016} and in the $Y$-band reported in the tenth data release of the Large Area Survey (LAS) of the UKIRT Infrared Deep Sky Survey \citep[UKIDSS;][]{Lawrence07}.  

From the afterglow imaging it was realized that the total observed flux was only minimally brighter than that of the host galaxy during our follow-up observations at $22.37$ hr post-burst. In the analysis throughout the paper we therefore only consider the archival photometric data for the modelling of the galaxy hosting GRB\,160804A (see specifically Sect.~\ref{ssec:sed}). While we do not have images that can clearly resolve the host galaxy, we find from the TNG images that the host is compact, with no apparent structure. We measure a seeing-corrected full-width-at-half-maximum (FWHM) of $1\farcs4 \times 0\farcs7$ ($0\farcs2$ error), which corresponds to a physical size of $10.5 \times 5.2$\,kpc (1.5\,kpc error) at $z=0.737$.

\begin{table} %[!h]
	\centering
	\begin{minipage}{\columnwidth}
		\centering
		\caption{Photometry of the GRB\,160804A afterglow and host galaxy. The magnitudes reported here are all in the AB magnitude system and have not been corrected for the expected Galactic extinction of $E(B-V)=0.023$ \citep{Schlafly11}. \textbf{References.} $^1$\citet{Malesani16}; $^2$\citet{Bolmer16}; $^3$\citet{Watson16}; $^4$This work. \label{tab:photdata}}
		\begin{tabular}{cccc}
			\noalign{\smallskip} \hline \hline \noalign{\smallskip}
			Telescope/ & Filter  & UT mid time &  Magnitude \\
			instrument &  &  & (AB) \\
			\noalign{\smallskip}\hline \noalign{\smallskip}
			TNG/DOLoRes & $r$ & 2016-08-04 21:22 & $20.93 \pm 0.03^{1,4}$ \\
			TNG/DOLoRes & $z$ & 2016-08-04 21:22 & $21.07 \pm 0.05^1$ \\
			MPG/GROND & $g'$ & 2016-08-05 00:15 & $21.50 \pm 0.03^2$ \\
			MPG/GROND & $r'$ & 2016-08-05 00:15 & $21.21 \pm 0.03^2$ \\
			MPG/GROND & $i'$ & 2016-08-05 00:15 & $20.85 \pm 0.03^2$ \\
			MPG/GROND & $z'$ & 2016-08-05 00:15 & $20.66 \pm 0.03^2$ \\
			MPG/GROND & $J$ & 2016-08-05 00:15 & $20.20 \pm 0.07^2$ \\
			MPG/GROND & $H$ & 2016-08-05 00:15 & $19.88 \pm 0.07^2$ \\
			MPG/GROND & $K$ & 2016-08-05 00:15 & $19.69 \pm 0.12^2$ \\
			HJT/RATIR & $r$ & 2016-08-05 05:09 & $21.29 \pm 0.02^3$ \\
			HJT/RATIR & $i$ & 2016-08-05 05:09 & $20.90 \pm 0.02^3$ \\
			HJT/RATIR & $Z$ & 2016-08-05 05:09 & $21.12 \pm 0.19^3$ \\
			HJT/RATIR & $Y$ & 2016-08-05 05:09 & $21.04 \pm 0.29^3$ \\
			HJT/RATIR & $J$ & 2016-08-05 05:09 & $20.52 \pm 0.23^3$\\
			HJT/RATIR & $H$ & 2016-08-05 05:09 & $> 20.51^3$\\
			TNG/DOLoRes & $r$ & 2016-08-11 21:15 & $21.07\pm 0.05^4$ \\
			\noalign{\smallskip}\hline \noalign{\smallskip}
			SDSS-DR13 & $u$ & Archival data & $22.04\pm0.38$\\
			SDSS-DR13 & $g$ & Archival data & $21.71\pm0.11$ \\
			SDSS-DR13 & $r$ & Archival data & $21.23\pm0.11$ \\
			SDSS-DR13 & $i$ & Archival data & $20.67\pm0.09$ \\
			SDSS-DR13 & $z$ & Archival data & $20.74\pm0.35$  \\
			UKIDSS-DR10 & $Y$ & Archival data &  $21.08\pm 0.21$ \\
			\noalign{\smallskip} \hline \noalign{\smallskip}
		\end{tabular}
		\centering
	\end{minipage}
\end{table}

\begin{figure*} % [ht!]
	\centering
	\epsfig{file=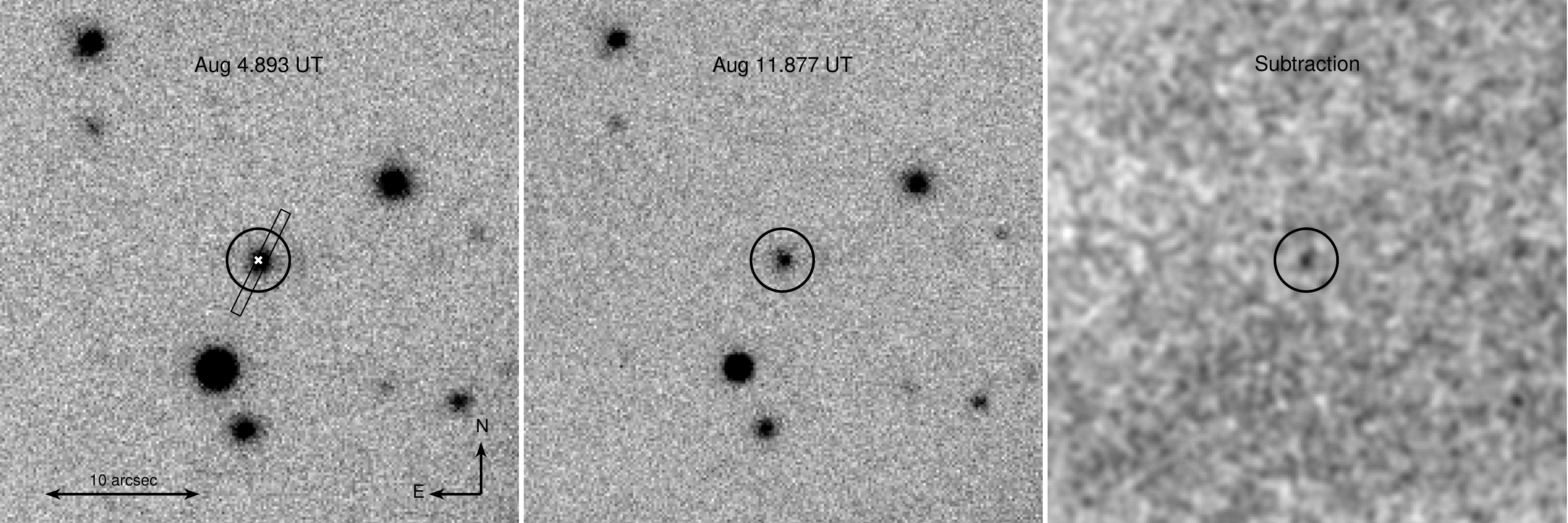,width=1.0\textwidth}
	\caption{Imaging of the afterglow and host of GRB\,160804A in the TNG $r$-band. 
		The left and middle panel show the first and second epochs, while in the right panel we show the results of image subtraction. A faint, but significant residual is detected at the host position in the TNG images obtained close in time to the X-shooter spectrum. The large black circle is centered on the residual position in all three images. The centroid of the residual is marked with a white cross in the first epoch imaging, where a schematic of the position of the slit used for the VLT/X-shooter observations is shown as well. \label{fig:imgsub}}
\end{figure*}

To get a better estimate of the contribution of the afterglow to the observed total flux in the X-shooter data we obtained a second epoch $r$-band observation with TNG (at $\Delta t = 7.8$ d post-burst). The first epoch was obtained close in time with the X-shooter spectrum ($\Delta t_{\mathrm{TNG-XS}} = 1.55$ hr), so that we can get an estimate of the upper limit for the afterglow contribution. After image subtraction of the two TNG $r$-band epochs (see Fig.~\ref{fig:imgsub}) we detect a faint residual at the host position with a brightness of $r=23.53\pm 0.14$ mag. Using the archival $r$-band imaging data from the SDSS we estimate that only a marginal contribution from the afterglow (around 15\%) was present at the time of the first TNG observation. This then implies that the X-shooter spectrum is host-dominated and that the strong absorption features are intrinsic to the host galaxy continuum spectrum. In Fig.~\ref{fig:imgsub} we also mark the centroid of the residual in the first epoch imaging with a white cross to show the relative distance of the afterglow to the host galaxy. We measure a projected relative distance of $0\farcs25 \pm 0\farcs1$ (i.e., $1.9 \pm 0.8$ kpc), well within the FWHM of the host galaxy.

Due to the peculiarity of the physical properties of the galaxy examined here compared to typical GRB hosts, we investigate the chance association probability of the observed galaxy to be an unrelated foreground galaxy. First, we downloaded and inspected the images taken by \textit{Swift}/UVOT \citep[e.g.,][GCN 19764]{Breeveld16}. The afterglow is faintly, but significantly, detected in the bluest filter (UVW2), which sets a (conservative) upper limit $z \la 1.7$. Then, our X-shooter spectrum does not reveal any emission (nor absorption) features at redshifts higher than $0.737$. While a very faint host galaxy cannot be excluded, our experience shows that we can normally detect emission lines with X-shooter up to $z \sim 2$ and beyond \citep[e.g.,][]{Kruehler12}. Another argument for the association between GRB\,160804A and the host comes from the low chance association probability to lie so close in projection to a bright galaxy: using Eq. 3 from \cite{Bloom02} we find a chance association probability of $9 \times 10^{-5}$. These arguments make it overall very likely that GRB\,160804A exploded in the $z=0.737$ galaxy that we have identified.

\section{Host galaxy properties in emission} \label{sec:hostem}

\subsection{Emission line analysis}

\begin{figure} %[ht!]
	\centering
	\epsfig{file=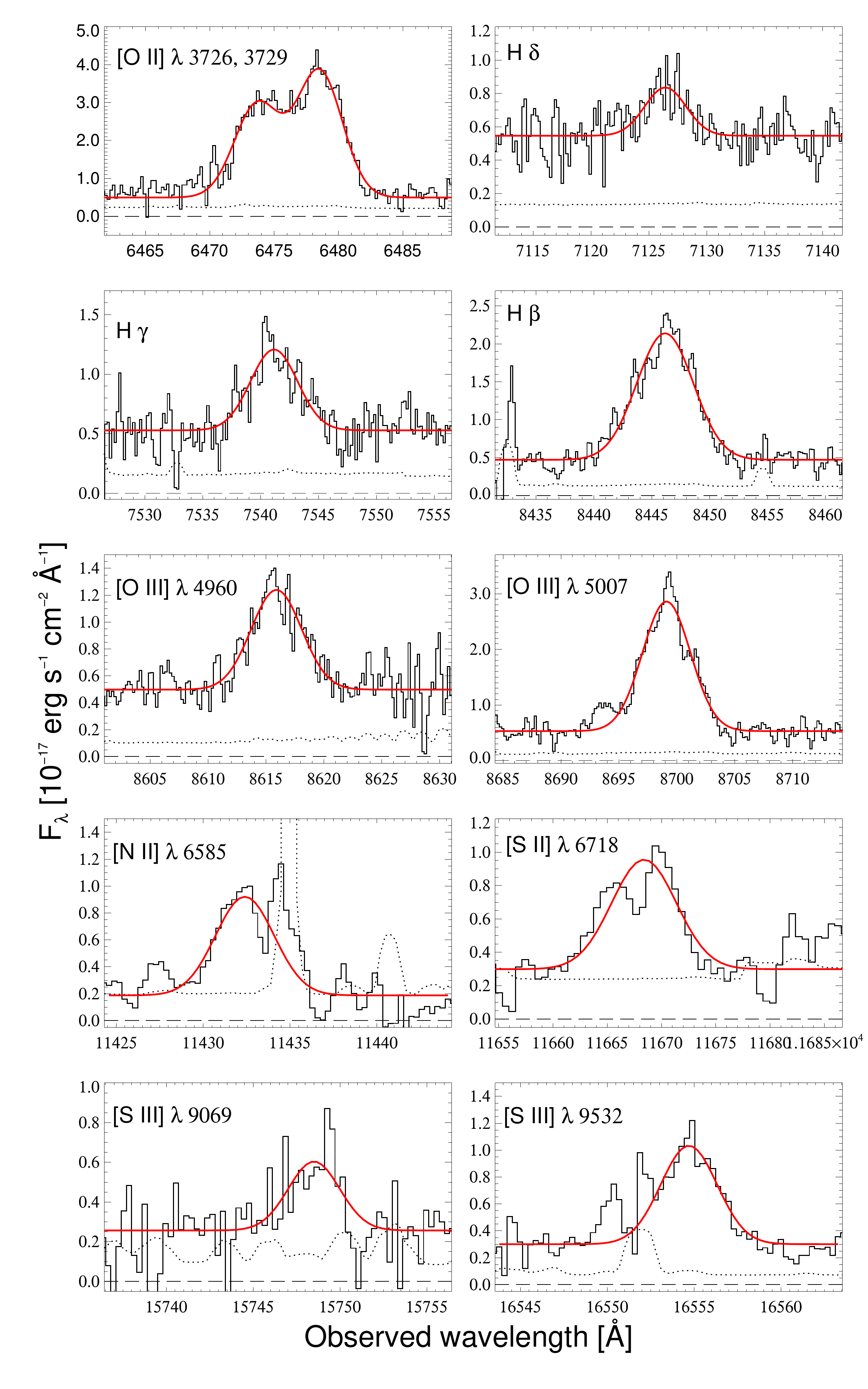,width=\columnwidth}
	\caption{Emission lines extracted from the X-shooter spectrum. In each panel the observed spectrum is shown by the solid, black line, the error spectrum by the dotted line and the best-fit Gaussian function by the solid, red line. The plotting region contains the continuum windows on either side of the line. \label{fig:emlines}}
\end{figure}

We detect several emission lines in the X-shooter spectrum, also reported by \cite{Xu16}. The measured transitions include the [O\,\textsc{ii}]\,$\lambda\lambda$\,3726,3729 doublet, the [O\,\textsc{iii}]\,$\lambda$\,4960 and [O\,\textsc{iii}]\,$\lambda$\,5007 lines, [N\,\textsc{ii}]\,$\lambda$\,6585, [S\,\textsc{ii}]\,$\lambda$\,6718, [S\,\textsc{iii}]\,$\lambda$\,9069, [S\,\textsc{iii}]\,$\lambda$\,9532 and the four Balmer lines, H$\alpha$, H$\beta$, H$\gamma$ and H$\delta$. To extract the line fluxes we fitted a Gaussian to each line with the continuum measured in small regions around the centroid of the fit ($\pm 30$ \AA), free of telluric- and sky-lines. From the fit we then determined the FWHM of each line. To correct for the instrumental broadening, $\Delta V$ (km s$^{-1}$), we subtracted it quadratically from the fitted FWHM given as $\mathrm{FWHM_{corr}} = \sqrt{\mathrm{FWHM_{obs}^2} - \Delta V^2}$. Based on the fits we measured a systemic emission line redshift of $z=0.73694 \pm 0.00003$ from the observed [O\,\textsc{ii}], [O\,\textsc{iii}] and Balmer lines. The results of the measured line fluxes and line widths are listed in Table~\ref{tab:emlines} and shown in Fig.~\ref{fig:emlines} with the best-fit Gaussian function, except for H$\alpha$ which will be studied in detail individually in Sect.~\ref{ssec:rot}. 

\begin{table*}%[!h]
	\centering
	\begin{minipage}{1.0\textwidth}
		\centering
		\caption{Extracted emission-line fluxes from the best-fit Gaussian functions. The line flux reported for H$\alpha$ is from a numerical integration of the line profile. \label{tab:emlines}}
		\begin{tabular}{lcccl}
			\noalign{\smallskip} \hline \hline \noalign{\smallskip}
			Transition & Line flux  & FWHM & Redshift & Notes \\
			& ($10^{-17}$ erg cm$^{-2}$ s$^{-1}$) &(km s$^{-1}$)  & & \\
			\noalign{\smallskip}\hline \noalign{\smallskip}
			\lbrack O\,\textsc{ii}\rbrack\,$\lambda$\,3726,3729 & $35.50\pm 0.15$ & $179.4 \pm 8.0$ & 0.73693 & Double-Gaussian \\
			H$\delta$\,$\lambda$\,4103 &  $1.13 \pm 0.05$ & $175.4 \pm 33.3$  & 0.73694  & \\
			H$\gamma$\,$\lambda$\,4342 & $3.47\pm 0.05$ & $188.0 \pm 16.9$ & 0.73692& \\
			H$\beta$\,$\lambda$\,4863 & $10.11\pm 0.04$ & $202.3 \pm 5.4$ & 0.73693 &\\
			\lbrack O\,\textsc{iii}\rbrack\,$\lambda$\,4960 & $3.99\pm 0.04$  & $176.7 \pm 10.2$ & 0.73697 &\\
			\lbrack O\,\textsc{iii}\rbrack\,$\lambda$\,5008 & $12.18\pm 0.04$ & $165.9 \pm 3.8$ & 0.73696 &  \\
			H$\alpha$\,$\lambda$\,6565 & $31.25\pm 0.12$ & $\cdots$ & 0.73704 & Integrated fit  \\ %103.4pm3.6
			\lbrack N\,\textsc{ii}\rbrack\,$\lambda$\,6585 & $5.17\pm 0.10$ & $205.4 \pm 40.5$ & 0.73634 &  \\
			\lbrack S\,\textsc{ii}\rbrack\,$\lambda$\,6718 &  $5.86\pm 0.11$ & $176.0 \pm 40.2$  & 0.73677 &  \\
			\lbrack S\,\textsc{iii}\rbrack\,$\lambda$\,9069 & $2.57\pm 0.06$ & $127.3 \pm 29.0$ & 0.73746 &  \\
			\lbrack S\,\textsc{iii}\rbrack\,$\lambda$\,9532 & $8.83\pm 0.04$ & $133.0 \pm 11.1$  & 0.73707 &  \\
			\noalign{\smallskip}\hline \noalign{\smallskip}
		\end{tabular}
		\centering
	\end{minipage}
\end{table*}

\subsection{Dust attenuation and star-formation rate}

The observed line strengths are affected by the amount of dust in the host galaxy, the dust attenuation, $A_V$, of the H\,\textsc{ii} regions. To correct for this effect we estimate the reddening of the system given the physical conditions of the emitting region. The ratio of the two Balmer lines, H$\alpha$ and H$\beta$ (known as the Balmer decrement), should intrinsically be $r_{\mathrm{int}}=$ H$\alpha$/H$\beta = 2.87$ \citep{Osterbrock89}, assuming a case B recombination and an electron temperature of $T_e = 10^4$ K and a density of $n_e = 10^2 - 10^4$ cm$^{-3}$ \citep[but see e.g.][for examples of GRB hosts in which these assumptions do not hold]{Wiersema11}. We measure a ratio of $r_{\mathrm{obs}}=$ H$\alpha$/H$\beta = 3.09\pm 0.02$, which we use to estimate the attenuation as
\begin{equation}
E(B-V) = \frac{2.5 \log(r_{\mathrm{obs}}/r_{\mathrm{int}}) }{k(\mathrm{H}\beta) - k(\mathrm{H}\alpha)} = 0.06 \pm 0.01~,
\end{equation}
where $k$ denotes the reddening law evaluated at the given wavelengths. The error reported here is only derived from the statistical errors of the line fluxes. When assuming the reddening law of \cite{Calzetti00}, with a total-to-selective $V$-band extinction parameter of $R_V=4.05$ \citep[common for star-forming galaxies, e.g.][]{Calzetti94,Calzetti00,Fischera03}, this equals a visual attenuation of $A_V=E(B-V)\,R_V=0.24\pm 0.04$ measured for the H\,\textsc{ii} regions.

To compute the star-formation rate (SFR) of the host galaxy we first use the emission redshift to derive the luminosity distance, $d_L$. Then we correct the observed H$\alpha$ line flux for the dust extinction found from the Balmer decrement and find the dereddend line flux (in general) as
\begin{equation}
f_{\mathrm{dered}} = f_{\mathrm{obs}} \times 10^{0.4\,E(B-V)\,k(\lambda)}~,
\end{equation}
which for H$\alpha$ equals a dereddened line luminosity of $L_{\mathrm{H}\alpha,\mathrm{dered}}=(9.59\pm 0.04)\times 10^{41}$ erg s$^{-1}$. Converting the luminosity into a SFR using the relation of \cite{Kennicutt98} yields
\begin{equation}
\mathrm{SFR}_{\mathrm{H}\alpha} = 7.9\,\times 10^{-42}\,L_{\mathrm{H}\alpha,\mathrm{dered}}~/~1.7 =4.46\pm 0.02~M_{\odot}~\mathrm{yr}^{-1}~,
\end{equation}
where the relation is given for a Salpeter IMF, which we have converted to that of \cite{Chabrier03}. Due to slit losses the true H$\alpha$ flux is underestimated so that the computed SFR$_{\mathrm{H}\alpha}$ should be interpreted as a lower limit. We estimate the slit loss by normalizing the observed spectrum to the $r$-band magnitude from the archival SDSS data and find that the observed flux should be increased by approximately 50\%, which should be propagated into the measurement of the SFR. Again, the errors reported here are only derived from the statistical errors of the line fluxes and do not include the scatter in the Kennicutt relation. 

We note that when measuring the line fluxes and from these deriving the dust content, $E(B-V)$, we do not include the effect of Balmer stellar absorption. \cite{Wiersema11} found that the majority of GRB hosts with high S/N spectra show these absorption features, however, we do not detect any indication of an absorption trough around the emission lines (see specifically their Fig.~5) and therefore do not include this effect in the line fits.

\subsection{Metallicity}

To infer the gas-phase metallicity of the host galaxy we use the strong-line ratios \citep[see][for a review]{Kewley08} of the dust-corrected transitions measured from the spectrum. Specifically, we calculate the metallicity using the $R_{23}$ calibration relating the dust-corrected line fluxes of [O\,\textsc{ii}], [O\,\textsc{iii}] and H$\beta$. This strong-line diagnostic, however, is double-valued for which an upper and lower branch exists. This degeneracy can be broken via the line ratio of [N\,\textsc{ii}]/[O\,\textsc{ii}], with values above $\mathrm{[N\,\textsc{ii}]/[O\,\textsc{ii}]} \approx -1.2$ indicative of the upper branch region. We measure $\mathrm{[N\,\textsc{ii}]/[O\,\textsc{ii}]} = -0.90\pm 0.01$ clearly validating this solution. Using the $R_{23}$ calibration of \cite{Zaritsky94} this yields a relative oxygen abundance of $12 + \log(\mathrm{O/H}) = 8.74\pm 0.12$. The error on the derivation includes the scatter in the relations listed in \cite{Kewley08}. Assuming a solar relative oxygen abundance of $12+\log(\mathrm{O/H})=8.69$ \citep{Asplund09}, this correponds to a metallicity consistent with solar ($Z = 1.12^{+0.36}_{-0.27}\,Z_{\odot}$). Relying instead on the O3N2 calibration \citep{Kewley08} we measure a relative oxygen abundance of $12 + \log(\mathrm{O/H}) = 8.64\pm 0.25$, consistent with the $R_{23}$ emission-line diagnostic. Compared to the compilation of GRB host galaxies observed with X-shooter \citep{Kruehler15}, this particular host is among the top 20\% most metal-rich hosts at redshifts below one. 

We note that it is also possible to derive the electron density and temperature directly from the observed line ratios, which would yield a more precise estimate of the metallicity. However, since we do not detect e.g. [S\,\textsc{ii}]\,$\lambda$\,6731 and the auroral line [O\,\textsc{iii}]\,$\lambda$\,4364 we are only able to put poorly constraining limits on the metallicity \citep[see e.g.][for a detailed example]{Wiersema07}. Furthermore, the metallicities derived for the comparison sample described in Sect.~\ref{fig:mzrel} are all based on typical metallicity calibrations of strong line ratios, so to compare we adopt the metallicity derived from the $R_{23}$ calibration.

While GRBs were initially predicted to occur in very metal-poor ($Z/Z_{\odot} < 0.2 - 0.3$) environments \citep{Yoon06,Woosley06b}, they are commonly observed in galaxies with higher oxygen abundances. This apparent contradiction can be resolved if the host galaxies have significant internal chemical heterogeneity, however, the observed metallicity gradients appear to be small \citep[see e.g.][]{Christensen08,Thoene14,Kruehler17,Izzo17}. For comparison, \cite{Niino11} predicts that over 10\% of GRB host galaxies are expected to have oxygen abundances of $12+\log(\mathrm{O/H})>8.8$, assuming the same internal dispersion of gas-phase metallicity to that observed in the Milky Way (which is indeed not a good assumption for GRB hosts), even if the GRB progenitors are metal-poor. Significant metallicity dispersion was also found to be a viable explanation for the origin of long-duration GRBs in metal-rich host galaxies from the Illustris simulation \citep{Bignone17}. In our case, however, the explosion site of the GRB was located well within the FWHM of the host galaxy light. The inferred gas-phase metallicity is therefore likely representative of the physical conditions in the local environment of the GRB.

\subsection{Rotation velocity} \label{ssec:rot}

\begin{figure} %[ht!]
	\centering
	\epsfig{file=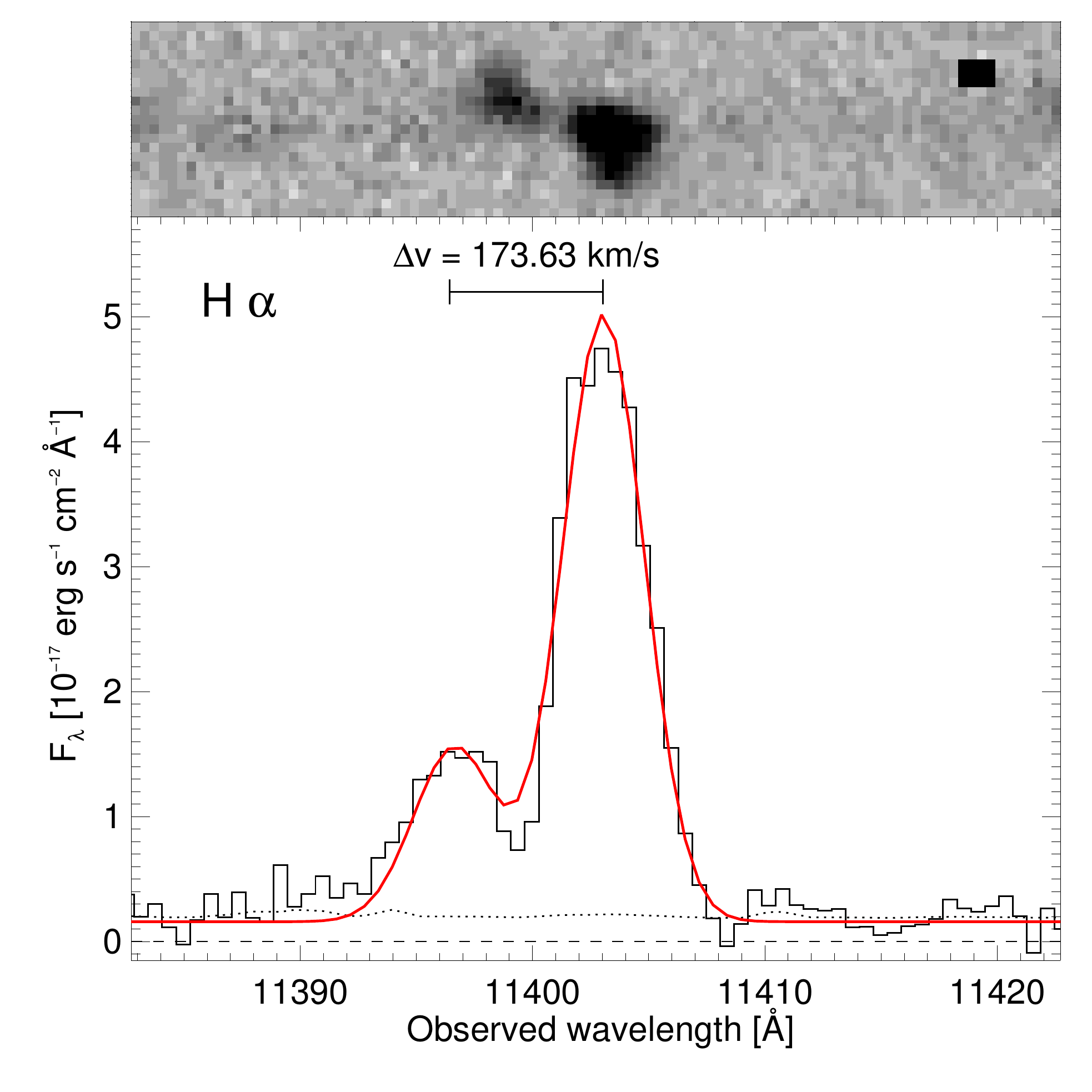,width=\columnwidth}
	\caption{The H$\alpha$ emission line extracted from the X-shooter spectrum. The top panel shows the observed 2D spectrum and the panel below shows the extracted 1D spectrum (black solid line) and the error spectrum (black dotted line). Overplotted is the best-fit double Gaussian function shown as the red solid line. The difference in velocity space of the two peaks is indicated at the top of the bottom panel. \label{fig:ha}}
\end{figure}

Only in a small number of cases, spatially-resolved analyses of GRB host galaxies have been possible due to their low redshift \citep[see e.g.][]{Christensen08,Thoene08,Levesque10a,Levesque11,Starling11,Thoene14,Michalowski16,Kruehler17,Tanga17}. In our case, while not resolved in imaging, we detect a spatially resolved line profile of H$\alpha$ (Fig.~\ref{fig:ha}) from which we can determine the rotational velocity of the system. As indicated in the figure we measure a difference of the two peaks in velocity space of $\Delta v = 173.63\pm 0.22$ km s$^{-1}$, i.e. a rotational velocity of $\Delta V_{\mathrm{rot}} = 86.82\pm 0.11$ km s$^{-1}$. Unfortunately, since we do not know the size and inclination of the host galaxy we can not determine a stellar mass based on a simple Tully-Fisher relation (but see Sects. \ref{ssec:sed} and \ref{ssec:massmet}). Here we have assumed that the blueshifted, weaker component is representing the spatial rotation of the galaxy. Another plausible scenario is that the weak emission component is emitted from a large-scale outflow or from a secondary galaxy merging with the primary (see Sect.~\ref{ssec:outfl}). The total line flux of H$\alpha$ is determined by simply integrating the line profile and is given in Table~\ref{tab:emlines}. A double-component Gaussian function is a good approximation to the data as well, and is overplotted in red in the figure. 

\begin{figure} %[ht!]
	\centering
	\epsfig{file=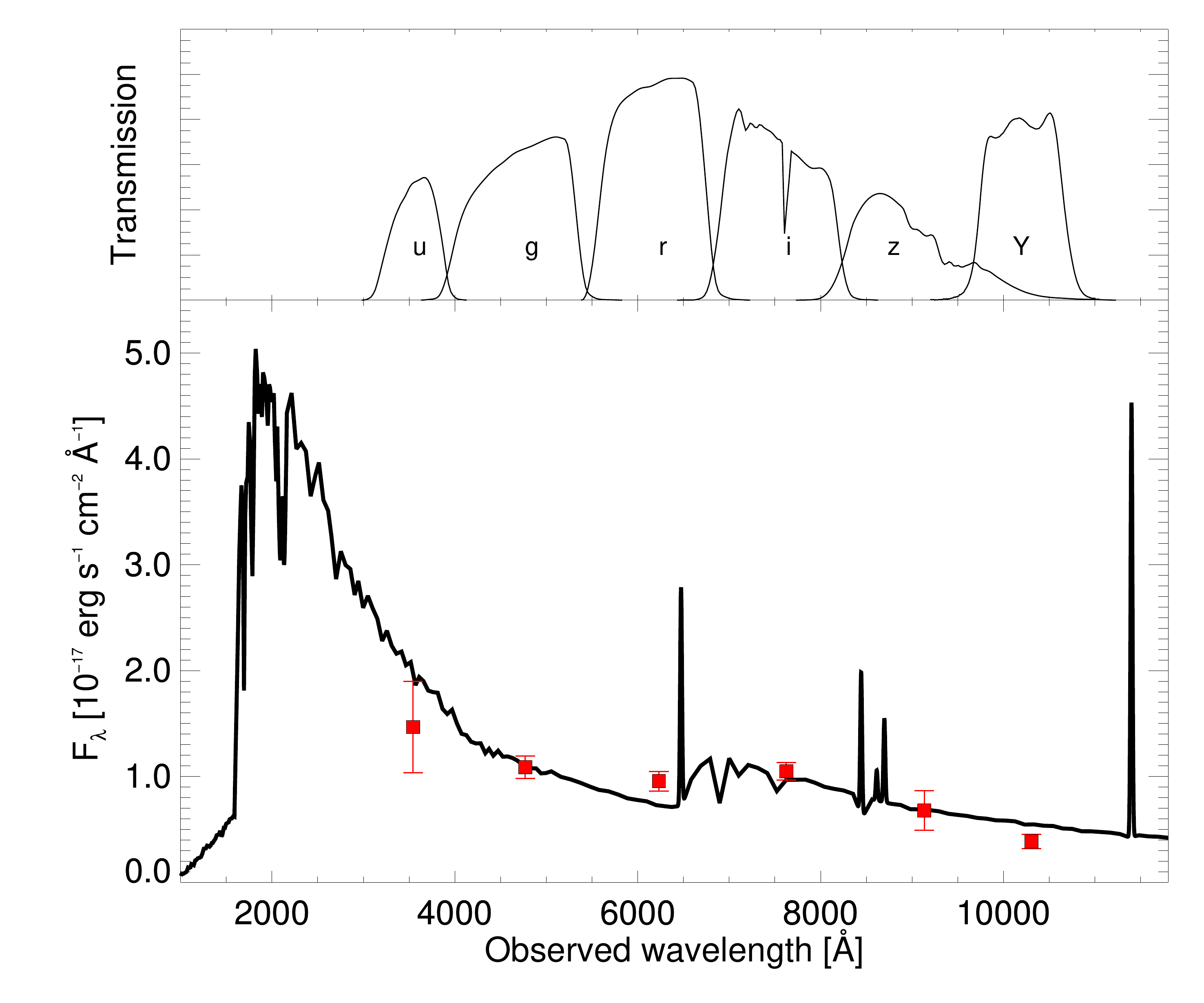,width=\columnwidth}
	\caption{SED of the host galaxy of GRB\,160804A. Overplotted is the best-fit stellar population synthesis model of a galaxy at a fixed redshift of $z=0.737$, reddened by $E(B-V)=0.01$ mag following the extinction curve of Calzetti et al. (2000). Also shown in the top panel are the optical SDSS ($ugriz$) and the near-infrared UKIDSS $Y$-band filter transmission curves for the corresponding six photometric points. \label{fig:sedfit}}
\end{figure}

\subsection{Broadband SED fit} \label{ssec:sed}

Fitting the six archival broad-band magnitudes reported in Table~\ref{tab:photdata} in LePhare\footnote{\url{http://www.cfht.hawaii.edu/~arnouts/LEPHARE/lephare.html}} \citep{Arnouts99,Ilbert06}, corrected for the foreground Galactic extinction, yields the best-fit stellar population synthesis model shown in Fig.~\ref{fig:sedfit} and the following host galaxy parameters: attenuation $E(B-V)=0.01^{+0.04}_{-0.01}$, star formation rate $\mathrm{(SFR)} = 5.57\pm 4.40~M_{\odot}$ yr$^{-1}$, stellar mass $\log (M_*/M_{\odot})= 9.80\pm 0.07$ and a stellar population of $\mathrm{age} = 286\pm 70$ Myr. We obtained the best fit SED by fixing the redshift to $z=0.737$ and used a grid of stellar evolution models with varying and exponentially decreasing star formation time scales, age of stellar population and extinction assuming the models from \cite{Bruzual03} based on an IMF from \cite{Chabrier03} and a Calzetti extinction curve \citep{Calzetti00}. The SFR from the best-fit SED is consistent with that estimated from the luminosity of the H$\alpha$ emission line and the model contains little to no dust ($E(B-V)< 0.05$) consistent with that inferred from the Balmer decrement. 

\begin{figure*} %[ht!]
	\centering
	\epsfig{file=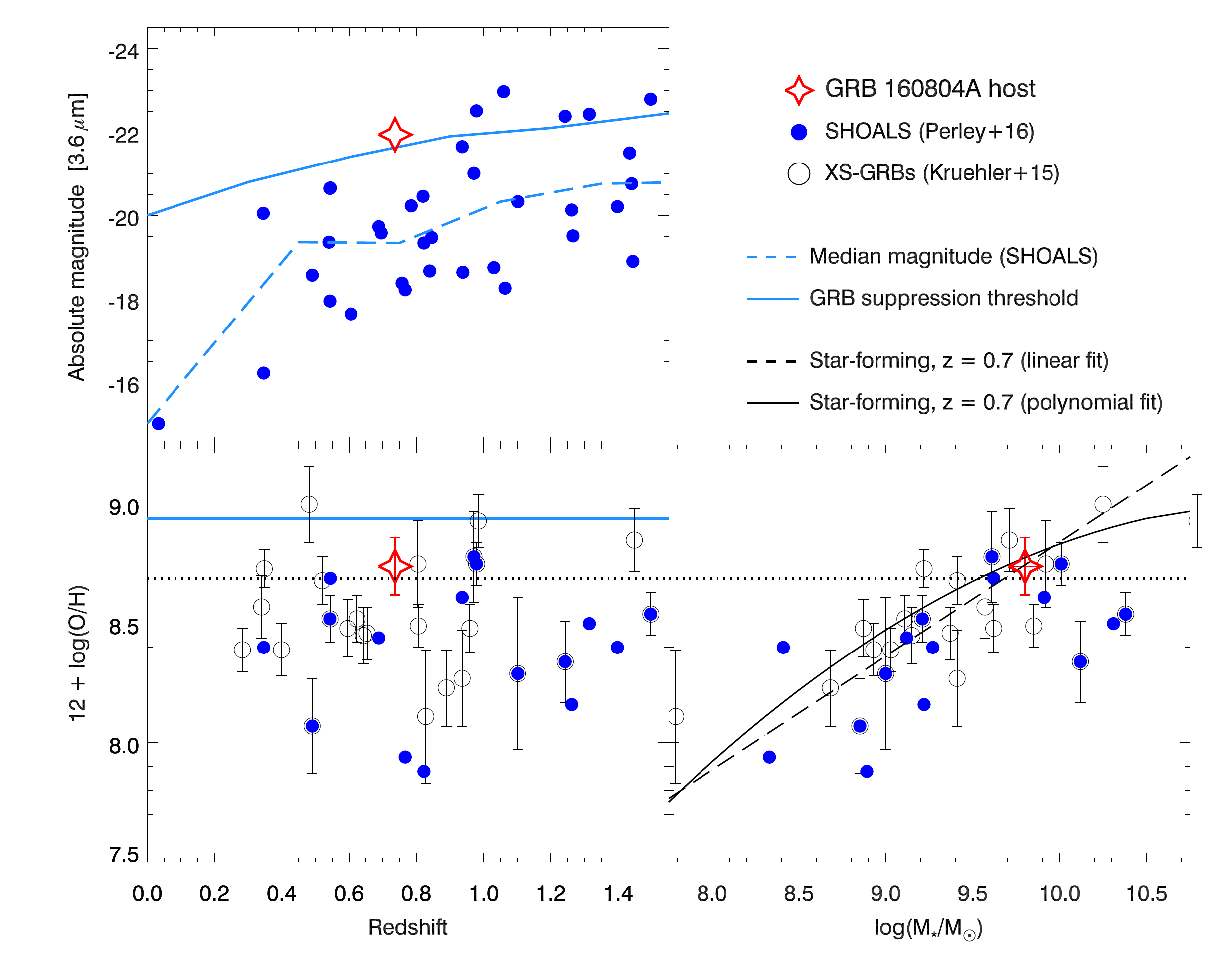,width=0.9\textwidth}
	\caption{Comparison of the host galaxy properties of GRB\,160804A (red star symbol) to other GRB hosts at $z$ < 1.5. We assume the oxygen abundance derived from the $R_{23}$ calibration in this figure. The blue dots denote sources in the unbiased GRB host galaxy sample, SHOALS, and the black circles show sources in the compiled sample of GRB hosts observed with X-shooter (see text). Only a subset of the sources in the SHOALS sample have emission-line metallicity measurements and computed stellar masses. In the upper panel the median magnitude of the SHOALS GRB host population (dashed blue) and the metallicity threshold derived by Perley et al. (2016b) converted to luminosity (solid blue) is also shown. In both the two lower panels, solar metallicity is marked as the dotted line. In the lower left panel the metallicity threshold is again shown as the solid blue line at $12+\log$(O/H) = 8.94. In the lower right panel the mass-metallicity relation of typical luminosity detected star-forming galaxies at $z\approx 0.7$ is overplotted as well. The host of GRB\,160804A is clearly among the most luminous, metal-rich and massive GRB host galaxies at these redshifts. \label{fig:mzrel}}
\end{figure*}

\subsection{Near-infrared luminosity} \label{ssec:lum}

It was recently shown by \cite{Perley16a,Perley16b}, based on the observationally unbiased \textit{Swift} GRB host galaxy legacy survey (SHOALS) that dusty ($A_V > 1$ mag) GRBs almost exclusively occur in the most infrared ($3.6\,\mu$m) luminous host galaxies. Furthermore, they derived an envelope for strong GRB suppression in metal-rich hosts, which acts as a soft upper limit. A similar conclusion was reached by \cite{Vergani15,Vergani17} based on the {\it Swift}/BAT6 sample of long-duration GRB hosts from which they were able to quantify a mild metallicity threshold of approximately $0.7\,Z_{\odot}$, above which GRB production is significantly suppressed. This metallicity suppression is argued to be the cause of the fact that GRB hosts appear to be significantly underluminous as a population, especially at redshifts below $z<1.5$, compared to typical luminosity-selected star-forming galaxies \citep{Perley16b,Vergani17}.

While we can not securely establish the extinction in the line of sight to the afterglow due to the small afterglow contribution at the time of observation, we have shown that the host galaxy does not appear to contain significant amounts of dust \citep[examples of low $A_V$ in the galaxy hosting GRBs with dusty afterglows are known though, see e.g.][]{Kruehler11}. 
In Fig.~\ref{fig:mzrel} (upper left panel) we compare the absolute magnitude of the host of GRB\,160804A to the GRB hosts in the SHOALS sample. We compute the absolute magnitude in the 3.6 $\mu$m-band of \textit{Spitzer} as
\begin{equation}
M_{3.6/(1+z)} = m_{\mathrm{obs},3.6} - \mu(z) + 2.5\log_{10}(1+z)~, \label{eq:absmag}
\end{equation}
where $\mu(z)$ is the distance modules given by $\mu(z)~=~5\log_{10}(d_L/10~\mathrm{pc})$ and $d_L$ is again the luminosity distance at the given redshift in units of parsec. The apparent 3.6 $\mu$m magnitude, $m_{\mathrm{obs},3.6}$, was computed by integrating the best-fit SED over the wavelength coverage of the 3.6 $\mu$m \textit{Spitzer} transmission curve from which we then measure $m_{\mathrm{obs},3.6}=20.77$ mag. This yields an absolute magnitude of $M_{3.6/(1+z)}=-21.94$ mag which makes the galaxy hosting GRB\,160804A among the most luminous 5\% of the GRB host population studied in the SHOALS sample. 

Moreover, as mentioned above, this fairly dust-poor host galaxy is located in a region in infrared-luminosity-space that is otherwise almost exclusively populated by obscured, dusty GRBs \citep{Perley16b}. In fact, the host galaxy of GRB\,160804A is clearly not a typical GRB host in that sense, but instead resembles more the luminosity-selected field galaxies from e.g. the MODS survey of the GOODS-North field \citep{Kajisawa11} or the sources in the CANDELS survey of the UDS field \citep{Galametz13,Santini15}. See e.g. the comparison made by \cite{Perley16b} specifically in their Figs. 3 and 5. GRB hosts with similar properties as the galaxy hosting GRB\,160804A were predicted by \cite{Trenti15} to constitute a significant fraction of $z<2$ hosts, while observations suggest this group to be sparsely populated. The case of GRB\,160804A is thus a step further towards this picture, lifting the threshold of metal-rich, luminous GRB hosts at $z<1$.

\subsection{Mass-metallicity relations at $z\approx 0.7$} \label{ssec:massmet}

So far it appears that the galaxy hosting GRB\,160804A more closely resembles the general luminosity-selected star-forming population than that of GRB host galaxies. Having computed the SFR, metallicity and stellar mass, we can now investigate whether this is also true in terms of the mass-metallicity ($M_* - Z$) relation at $z\sim 0.7$ \citep[e.g.,][]{Savaglio05}. The overall issue is still under debate: several studies have found that GRB hosts generally fall below the standard $M_* - Z$ relation \citep{Stanek06,Kewley07,Levesque10b,Han10,Mannucci11,Graham13,Japelj16,Vergani17} whereas e.g. \cite{Arabsalmani17} find that the offset could also partially be explained by systematic effects. Since galaxies hosting GRBs are a subset of the general population of star-forming galaxies, they are expected to follow the same $M_* - Z$ relation, although typically located in the low-mass end. The intrinsic properties of GRB hosts, such as large outflows, higher specific star formation rates and star formation densities or higher gas fractions, however, could cause the observed offset to lower metallicities \citep{Hughes13,Arabsalmani17}.

The linear bisector best-fit of the $M_* - Z$ relation at $z\sim 0.7$, computed from 56 galaxies at $0.4 < z < 1.0$ from the Gemini Deep Deep Survey (GDDS) and Canada-France Redshift Survey (CFRS), was derived by \cite{Savaglio05} to be
\begin{equation}
12+\log(\mathrm{O/H}) = (0.478\pm 0.058)\log(M_{*}/M_{\odot})+(4.062\pm 0.579)~.
\end{equation}
Based on the measured emission-line oxygen abundance in our case, 12+$\log(\mathrm{O/H}) = 8.74\pm 0.12$ (from the $R_{23}$ calibration), this yields an expected stellar-mass of $\log(M_{*}/M_{\odot})=9.79\pm 0.05$ should it follow the same trend as the general star-forming galaxy population. This is perfectly consistent with that found from the best-fit SED ($\log(M_{*}^{\mathrm{SED}}/M_{\odot})=9.80\pm 0.07$).

We show the above linear relation and also the converted polynomial $M_* - Z$ relation for star-forming galaxies at $z\sim 0.7$ from \cite{Savaglio05} in Fig.~\ref{fig:mzrel} (lower right panel) and how they both intercept with the GRB\,160804A host galaxy in the $M_* - Z$ plane. We also compare these two general mass-metallicity relations to the sources in the unbiased SHOALS sample and the compilation of GRB hosts observed with X-shooter \citep{Kruehler15} in the redshift range $0.0 < z < 1.5$. Only a subset of the SHOALS host galaxies have emission-line metallicities \citep[only those that overlap with the sources in the samples by][]{Kruehler15,Japelj16,Vergani17} and stellar masses \citep[Kr\"uhler \& Schady, in preparation; see also][]{Arabsalmani17}. We note that the stellar masses derived from the $M_{3.6/(1+z)}$ luminosity in the SHOALS sample is typically 0.2 -- 0.3 dex larger than the stellar masses obtained by fitting the same host galaxy SED from multiple optical or NIR photometric data points.
This discrepancy was found by comparing the few sources in the SHOALS sample that overlaps with the hosts studied by Kr\"uhler \& Schady for which multi-wavelength photometry has been obtained. It is clear that the GRB hosts from the SHOALS sample generally fall below both of the two $M_* - Z$ relations of typical star-forming galaxies at similar redshifts, more so than those from the XS-GRB host sample (which could be dominated by significant selection bias). The average offset between the hosts in the SHOALS survey and the polynomial $M_* - Z$ relation for the general population of star-forming galaxies at $z=0.7$ is 0.2 dex. Also, only two out of nine ($\approx 20\%$) of the SHOALS sources have super-solar metallicities, whereas seven out of 18 ($\approx 40\%$) of the XS-GRBs have oxygen abundances higher than solar. This discrepancy is likely caused by the fact that the SHOALS sample is selected in an unbiased way, whereas the XS-GRB sample is merely a compilation of GRB hosts observed with VLT/X-shooter.

This establishes the galaxy hosting GRB\,160804A as one of the most luminous, massive and metal-rich GRB hosts at $z<1.5$. While the low dust attenuation is unusual for such a massive and metal-rich galaxy, comparing it to the sample of star-forming galaxies presented by \cite{Garn10} shows that, even though it falls below the general trend, it is still located within the scatter of attenuation as a function of SFR, metallicity and stellar mass. Furthermore, the properties of SFR, infrared luminosity, metallicity and stellar-mass closely follow the same prescriptions as for typical field galaxies in the same redshift range. Also, the host galaxies of the unbiased SHOALS sample with existing emission-line metallicities clearly fall below the general star-forming mass-metallicity relation at low-$z$'s, supporting previous claims.

\section{Host galaxy absorption line analysis} \label{sec:abs}

\subsection{Line identification}

%We detect a range of metal absorption lines in the X-shooter spectrum of GRB\,160804A, all located in the UVB arm (again, see Fig.~\ref{fig:uvbnorm}). 
%Due to the low-$z$ nature of the burst we can not detect Lyman-$\alpha$ since it is located bluewards of the atmospheric cut-off, so unfortunately it is not possible to determine an absorption line metallicity. Only a small number of GRBs are located at redshifts where both methods can be applied and compared (i.e. $1.8 < z_{\mathrm{GRB}} < 2.5$ for GRBs observed with VLT/X-shooter), but \cite{Friis15} find both the metallicity derived in absorption and emission of GRB\,121024A to be consistent, although the two methods do not necessarily probe the same regions of the galaxy hosting the GRB. See also \cite{Krogager13} for a similar study of the galaxy counterpart of a DLA towards a bright quasar.

All metal lines are identified in the UVB arm of the spectrum, where we detect the absorption features from singly-ionized iron: Fe\,\textsc{ii}~$\lambda$~2344,  Fe\,\textsc{ii}~$\lambda$~2374,  Fe\,\textsc{ii}~$\lambda$~2382, Fe\,\textsc{ii}~$\lambda$~2586, Fe\,\textsc{ii}~$\lambda$~2600 and magnesium: Mg\,\textsc{ii}~$\lambda$~2796, Mg\,\textsc{ii}~$\lambda$~2803 and Mg\,\textsc{i}~$\lambda$~2852. They all lie at a common redshift of $z=0.737$ and do therefore not belong to an intervening absorber. This indicates that we have a rare occurence of a GRB where the absorption lines of the galaxy can be detected even in absence (approximately 15\% of the total flux, recall Sect. \ref{ssec:ima}) of a strong underlying afterglow continuum.
We searched for any of the typical Fe\,\textsc{ii} fine-structure lines \citep{Christensen11} since these must be UV-pumped by a strong radiation field \citep{Prochaska06a,Vreeswijk07,Delia09}, which can only originate in the GRB afterglow. We do not detect any of the fine-structure lines (though we expect them to be weak in any case during the $\approx 22$ hr post-burst observations), further supporting that the spectrum and absorption lines are dominated by the host.

To determine the equivalent width ($W$) of the strong absorption lines, we fitted the continuum around each of the lines in regions that were free of contaminating absorption features and tellurics. We then summed over the absorption profile contained below the normalized flux level. The results are listed in Table~\ref{tab:ewabs}. All features are heavily saturated, partly due to the medium resolution of the spectra where high-resolution spectra of GRB afterglows \citep{Prochaska06b} have shown that strong metal absorption lines usually consist of a number of narrow components. We therefore can not derive a reliable column density based on Voigt profile fitting. Furthermore, there is evidence for a scenario where the absorption lines do not only probe the small-scale velocity components in the line of sight to the burst, but is rather a sum of the whole system as described below. This scenario would also be expected if the strong absorption lines are intrinsic to the host and not just a snapshot from the illumation by the afterglow emission.

\begin{table} %[!h]
	\centering
	\begin{minipage}{\columnwidth}
		\centering
		\caption{Measured rest-frame equivalent widths and the derived curve-of-growth column densities. The column densities derived for each transition are only to be taken as lower limits due to heavy line saturation. \label{tab:ewabs}}
		\begin{tabular}{lccccccc}
			\noalign{\smallskip} \hline \hline \noalign{\smallskip}
			Transition & $\lambda_{obs}$  & $W_{\mathrm{rest}}$ & $\log N_{\mathrm{CoG}}$ \\
			& (\AA) & (\AA) & (cm$^{-2}$)  \\
			\noalign{\smallskip}\hline \noalign{\smallskip}
			Fe\textsc{ ii}~$\lambda$~2344 & 4071.63 & $2.26\pm 0.69$ & $14.61\pm 0.12$ \\
			Fe\textsc{ ii}~$\lambda$~2374 & 4124.51 &$1.31\pm 0.55$ & $14.92\pm 0.15$  \\
			Fe\textsc{ ii}~$\lambda$~2382 & 4138.08 & $2.15\pm 0.55$ & $14.09\pm 0.10$ \\
			Fe\textsc{ ii}~$\lambda$~2586 & 4492.76 & $1.91\pm 0.62$ & $14.67\pm 0.12$  \\
			Fe\textsc{ ii}~$\lambda$~2600 & 4515.27 & $4.28\pm 0.70$ & $14.48\pm 0.07$  \\
			Mg\textsc{ ii}~$\lambda$~2796 &  4855.91 & $3.65\pm 0.58$ & $13.94\pm 0.06$  \\
			Mg\textsc{ ii}~$\lambda$~2803 & 4868.22 & $4.98\pm 0.56$ & $14.37\pm 0.05$  \\
			M\textsc{ i}~$\lambda$~2852 & 4954.84 & $1.90\pm 0.37$ &$13.16\pm 0.08$\\
			\noalign{\smallskip}\hline \noalign{\smallskip}
		\end{tabular}
		\centering
	\end{minipage}
\end{table}

\subsection{Column densities from the curve of growth}

Due to the heavy line saturation in our medium resolution spectrum we have to rely on a curve of growth (CoG) analysis to derive column densities for each element. The CoG directly relates the measured, rest-frame $W$ to the column density on the linear part of the CoG where the optical depth is low ($\tau_0 < 1$) and the lines are not saturated. This relation depends on the Doppler $b$ parameter on the flat part of the curve, where this value can typically be inferred from the CoG analysis \citep[see e.g.][for a detailed description]{Thoene10}.

It is preferred to analyze lines that are not heavily saturated, which is not possible in our case. Therefore, we only derive lower limits to the column densities for each ion given by
\begin{equation} \label{eq:cogN}
N = \frac{W}{\lambda}\frac{1.13\times 10^{20}\,\mathrm{cm}^{-2}}{f\,\lambda\, [\AA]^2}~,
\end{equation}
where $f$ is the oscillator strength of each transition, $W$ the equivalent width and $\lambda$ is the rest-wavelength in units of \AA~of each transition.
The values are reported in Table~\ref{tab:ewabs}.

%\begin{figure}% [!h]
%	\centering
%	\epsfig{file=CoG_all.pdf,width=\columnwidth}
%	\caption{Multi-ion single-component curve of growth fit for the eight detected absorption lines. The CoG is expressed in units of the column density, $N$, the oscillator strength of each transition, $f$, and the rest-frame wavelength of the transition. We show curves of growth for a range of effective Doppler parameter, $b$, values where $b=\infty$ is plotted as the solid line. Due to the heavy saturation of each line we are not able to securily derive column densities, and all should be considered as lower limits. \label{fig:cog}}
%\end{figure}

%One point is striking: the best-fit $b$ parameter is extremely large ($> 300$ km s$^{-1}$). We interpret the cause of this as being due to a sum of velocity components in the host. In practice, this is represented by the heavily saturated lines which are insensitive to the $b$ parameter. This then implies that all the column densities derived from the MISC-CoG analysis should only be considered as lower limits. In general, if an effective Doppler parameter of $b \gg 20$ km s$^{-1}$ is measured, \cite{Prochaska06b} showed that such strong lines actually consists of a range of velocity components, each with $b < 20$ km s$^{-1}$. We also note that the different ions should be treated in separate analyses as they might not probe the same regions of the galaxy. 

\subsection{Kinematics}

\begin{figure}% [!ht]
	\centering
	\epsfig{file=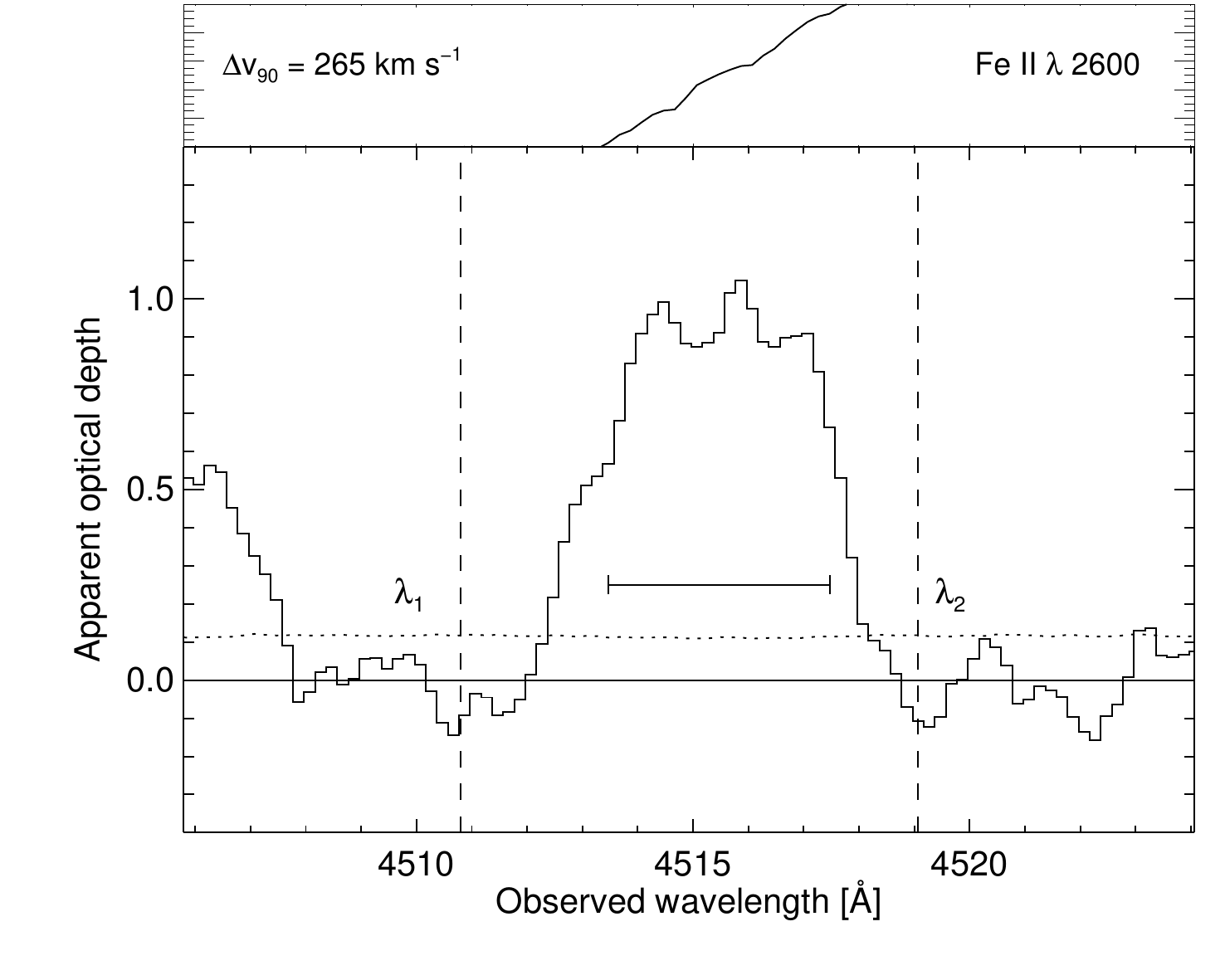,width=\columnwidth}
	\caption{The absorption line profile of the Fe\,\textsc{ii}\,$\lambda$\,2600 transition. The velocity width, $\Delta v_{90}$, is measured to be 265 km s$^{-1}$ and is shown by the solid line near the center of the profile (cumulative histogram shown in the upper panel). The dashed lines marked $\lambda_1$ and $\lambda_2$ denote the start and end wavelengths used to integrate the profile. The error spectrum is shown as the dotted line. \label{fig:v90}}
\end{figure}

The velocity dispersion, $\sigma$, of the H\,\textsc{ii} regions in the galaxy hosting the GRB can be estimated by the average velocity dispersions of the Balmer and [O\,\textsc{iii}] transitions. We calculate this by dividing the corrected, mean FWHM of these lines by $2\sqrt{2\ln2}$. From the nebular lines of the H\,\textsc{ii} regions we derive $\sigma = 78 \pm 17$ km s$^{-1}$. To compare the kinematics of the emission lines with that measured in absorption, we compute $\Delta v_{90}$, the velocity that contains 90\% of the area under the apparent optical depth spectrum, following the procedure of e.g. \cite{Prochaska97,Ledoux06,Fynbo10} and \cite{Arabsalmani17}, for all the absorption features, represented by the saturated Fe\,\textsc{ii}\,$\lambda$\,2600 line in Fig.~\ref{fig:v90}. We caution that since the absorption lines are saturated the width will tend to be overestimated \citep{Ledoux06}. The absorbing material is also not in the line-of-sight to a point source, as is typically the case for quasars and GRB afterglows. Instead, the absorption lines probe an integrated region of the whole galaxy so that the estimate of $\Delta v_{90}$ are not comparable to the values derived for typical GRB hosts \citep[e.g.][]{Arabsalmani15,Arabsalmani17}.

The velocity width is measured to be $\Delta v_{90}=265$ km s$^{-1}$. If the apparent optical depth of absorption profiles are identical to the emission line profiles, the two velocity widths should relate as $\Delta v_{90} = 3.29\sigma$ \citep[due to how they are defined, see e.g. the discussion in][]{Arabsalmani17}. Converting the measured emission-line velocity dispersion into $\Delta v_{90}$ following this definition yields $\Delta v_{90} = 260 \pm 60$ km s$^{-1}$, perfectly consistent with that measured directly from the absorption line profiles. This then supports the scenario that the absorption lines trace the same underlying velocity components of the host galaxy as the emission lines, compared to what is typically observed for the sightline probing only the neutral gas towards the GRB.

\begin{figure}% [!h]
	\centering
	\epsfig{file=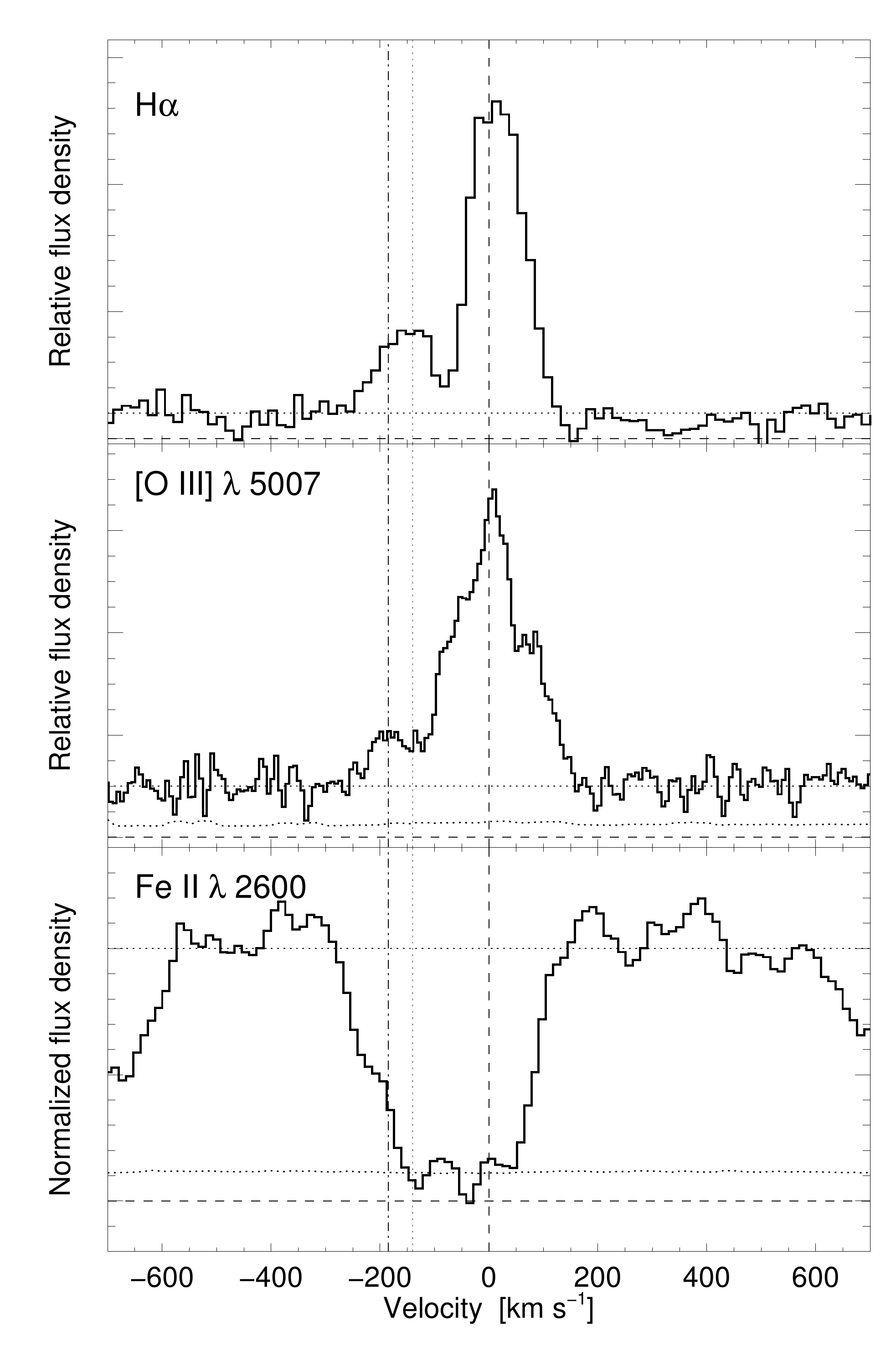,width=\columnwidth}
	\caption{The velocity profiles of the H$\alpha$ and [O\,\textsc{iii}]\,$\lambda$\,5007 emission lines (top and middle) and the normalized low-ionization absorption lines represented by Fe\,\textsc{ii}\,$\lambda$\,2600 (bottom). The respective error spectra are overplotted as the dotted lines. The zero-point of the velocity scale is set to the systemic redshift of $z=0.73696$ (dashed line) measured from the emission lines. The dotted and dash-dotted lines show the centroid of the two blueshifted components seen in emission from H$\alpha$ and [O\,\textsc{iii}], respectively. \label{fig:kin}}
\end{figure}

The centroids of the emission and absorption line profiles hold valuable information of the systemic redshift of the host galaxy as well. Specifically, the Balmer and [O\,\textsc{iii}] emission lines are thought to trace the H\,\textsc{ii} regions in the host, and thus the redshift of the star-forming component. In Fig.~\ref{fig:kin} we compare the H$\alpha$ and the [O\,\textsc{iii}]\,$\lambda$\,5007 emission lines to the low-ionization, satured Fe\,\textsc{ii}\,$\lambda$\,2600 absorption line. The absorption redshift is very similar to the systemic redshift. We note, however, that the shape of the absorption line profile appears to show similar resolved features as the emission lines but with a much stronger blue component. Similar line shapes of [O\,\textsc{iii}]\,$\lambda$\,5007 were also reported by \cite{Wiersema07} and \cite{Thoene07}, who also found shifts in the emission and absorption line centroids. The exact centroid of the absorption feature is hard to disentangle from the significant noise fluctuations which will also affect the measurement. While [S\,\textsc{ii}]\,$\lambda$\,6718 also appears to split into two components, none of the two match those observed in H$\alpha$ and [O\,\textsc{iii}]\,$\lambda$\,5007. It is likely that the observed double profile is simply an artifact of the poor signal-to-noise in the spectral region where this line is located.

\subsection{Evidence for a galactic-scale outflow} \label{ssec:outfl}

From a study of absorption features in a large sample of field-selected galaxies, \cite{Rubin14} found that outflows are ubiquitous in normal, star-forming galaxies at $z\approx 0.5$. Specifically, the SFRs of the galaxies were tightly correlated with the outflow equivalent widths of Fe\,\textsc{ii}\,$\lambda$\,2600 and Mg\,\textsc{ii}\,$\lambda$$\lambda$\,2796, implying that larger SFRs cause an increased ejection of the absorbing clouds. Since the absorption line profile is consistent with having an additional component at $\approx -200$\,km\,s$^{-1}$ relative to the systemic redshift based on the strongest component of the emission lines (Fig.~\ref{fig:kin}), there is evidence for a scenario where a galactic-scale outflow is being ejected from the galaxy hosting the GRB\,160804A. This is also a plausible explanation for the additional emission component seen in Fig.~\ref{fig:ha} for H$\alpha$. Furthermore, the line transitions Fe\,\textsc{ii}\,$\lambda$$\lambda$\,2586,\,2600 and Mg\,\textsc{ii}\,$\lambda$$\lambda$\,2796,\,2803 are consistent within the errors to represent a covering fraction of the absorbing material of approximately 100\%, which is rare even in the large sample analyzed by \cite{Rubin14}. The authors note that the expelled gas from the ejective stellar feedback process in these galaxies might be a viable source of material for the massive gas reservoirs observed as the circumgalactic medium around present day galaxies. The galaxy hosting GRB\,160804A could then be a case of a star-forming galaxy ejecting extreme amounts of cool gas, covering the full size of the galaxy it was expelled from.

Intriguingly, such large outflows (and the additional emission component seen in H$\alpha$ and tentatively in [O\,\textsc{iii}]) might also be indicative of a scenario where the observed host galaxy is in a late-stage merger \citep[just before fusion, see e.g.][]{Soto12}. The large stellar mass, star-formation rate and kinematics of the host galaxy studied here are consistent with that of the galaxy at $z=0.41$ examined by \cite{Peirani09}, who found via $N$-body simulations that such characteristics could be explained by two merged galaxies near coalescence \citep[see also][for a general study]{Hammer09}. This scenario would then explain the apparent discrepancy of this GRB host compared to others. Moreover, it is then possible that the secondary merged galaxy is actually metal-poor, in which the GRB explosion might have occurred.

\section{Conclusions} \label{sec:conc}

In this work we presented VLT/X-shooter spectroscopic follow-up observations of the \textit{Swift}-detected GRB\,160804A at $z=0.737$. We examined both the emission and absorption line properties of the galaxy hosting this burst in detail and modelled it with archival photometric data to complement the newly obtained spectroscopic data.

From the data analysis we found the galaxy to show strong, prominent absorption features from several metal lines, indicative of a high metallicity, even in the absence of a bright underlying continuum such as the typical scenarios where the afterglow outshines the galaxy hosting the GRB. From emission-line diagnostics we confirmed that the galaxy is indeed metal-rich, with a relative oxygen abundance consistent with solar ($12+\log(\mathrm{O/H}) = 8.74\pm 0.12$). Moreover, given the relative position of the explosion site of the GRB to the host galaxy there is tentative evidence that the GRB exploded in a metal-rich environment. The host appears to be dust-poor ($E(B-V) < 0.05$ mag), which is unusual for such large metallicities. Based on the line luminosity of H$\alpha$ we estimate the host to be forming stars at a rate of $4.46\pm 0.02~M_{\odot}~\mathrm{yr}^{-1}$ (the error reported here is only derived from the statistical errors of the line fluxes and does not include the scatter from the Kennicutt relation) and from the detected nebular lines we measure an average velocity dispersion of $78\pm 17$ km s$^{-1}$. Both the small amount of dust and the star-formation rate is reproduced by fitting stellar population synthesis models to the six broad-band photometric data points, where we also determine a stellar mass of $\log(M_*/M_{\odot})= 9.80\pm 0.07$.

An important piece to the puzzle of whether galaxies hosting GRBs are true tracers of star formation, or if they are biased toward lower metallicites, can be resolved by comparing an observationally unbiased sample of GRB hosts to the general population of star-forming galaxies. While this case clearly resembles more that of luminosity-selected field galaxies at $z\approx 0.7$ in terms of luminosity, mass and metallicity, GRB hosts are generally found to be sub-luminous and to fall below the mass-metallicity relation of main-sequence star-forming galaxies. Specifically, we have here shown that a subset of the host galaxies at $z<1.5$ in the observationally unbiased SHOALS survey with reported emission-line metallicities are on average 0.2 dex below the general mass-metallicity relation.

In addition to the observed nebular emission lines we found that the strong absorption lines present in the spectrum are representative of the sum of velocity components in the host galaxy. Whereas GRB afterglows reveal velocity components of discrete clouds along the line of sight, we find that the observed strong absorption features actually show a rare occurance of integrated host galaxy velocity components. These lines were furthermore detected even in the absence of a dominating afterglow continuum and must therefore be intrinsic to the host galaxy. Moreover, we found evidence for a scenario where the absorbing material is actually a galactic-scale outflow, covering $\approx 100\%$ of the galaxy it was expelled from. Large scale outflows are found in the majority of $z \approx 0.5$ star-forming galaxies, but the case presented here is one of the most extreme. We argue that since such outflows are produced by intense star formation events, a scenario where the host galaxy is actually a late-stage merger might be plausible. This scenario could also explain the discrepancy between the properties of this specific GRB host galaxy compared to others at similar redshifts.

\section*{Acknowledgements}
We would like to thank the anonymous referee for a constructive report provided in a timely manner.
KEH and PJ acknowledge support by a Project Grant (162948--051) from The Icelandic Research Fund. The research leading to these results has received funding from the European Research Council under the European Union's Seventh Framework Program (FP7/2007--2013)/ERC Grant agreement no. EGGS--278202. AdUP, CCT and ZC acknowledge support from the Spanish Ministry of Economy and Competitivity under grant number AYA 2014-58381-P. AdUP and CCT acknowledge support from Ramon y Cajal fellowships (RyC-2012-09975 and RyC-2012-09984). AdUP acknowledges support from a grant from the BBVA foundation for researchers and cultural creators. ZC acknowledges support from the Juan de la Cierva Incorporaci\'on fellowship IJCI-2014-21669 and from the Spanish research project AYA 2014-58381-P. 

%%%%%%%%%%%%%%%%%%%%%%%%%%%%%%%%%%%%%%%%%%%%%%%%%%



\bibliographystyle{mnras}
\bibliography{ref} % if your bibtex file is called example.bib

%\begin{thebibliography}{99}
%\end{thebibliography}

%%%%%%%%%%%%%%%%%%%%%%%%%%%%%%%%%%%%%%%%%%%%%%%%%%

%%%%%%%%%%%%%%%%% APPENDICES %%%%%%%%%%%%%%%%%%%%%

%\appendix

%%%%%%%%%%%%%%%%%%%%%%%%%%%%%%%%%%%%%%%%%%%%%%%%%%

% Don't change these lines
\bsp	% typesetting comment
\label{lastpage}
\end{document}